# Observations and Theoretical Calculations of 11-Year Cyclic Variations in Lower-Stratospheric Ozone Depletion and Cooling

Qing-Bin Lu

Department of Physics and Astronomy and Departments of Biology and Chemistry, University of Waterloo, 200 University Avenue West, Waterloo, Ontario, Canada (Email: qblu@uwaterloo.ca)

**Abstract:** Observations and quantitative understanding of spatio-temporary variations in lower-stratospheric ozone and temperature can provide fingerprints for the mechanisms of ozone depletion and play an important role in testing the impact of non-halogen greenhouse gases on the ozone layer in climate models. Here we report from ground-based ozonesonde and satellite-based measurements since the 1960s and 1979 respectively that both lower-stratospheric ozone and temperature display pronounced 11-year cyclic variations over Antarctica and mid-latitudes, while no apparent cyclic variations over the tropics. These observations were unexpected from the chemistry-climate models (CCMs) but predicted by the cosmic-ray-driven electron-induced-reaction (CRE) model of ozone depletion. Remarkably, no-parameter CRE theoretical calculations give the ozone loss vertical profile in perfect agreement with observations at the Antarctic Syowa station and excellently reproduce time-series variations of both lower-stratospheric ozone and temperature in all global regions. Furthermore, the large lower-stratospheric ozone depletion over the tropics in the 1980s and 1990s is also reproduced by CRE calculations. Moreover, CRE calculations exhibit complex phenomena in future trends of lower-stratospheric ozone and temperature, which are strongly affected by the future trends of cosmic-ray fluxes. The latter might even lead to almost no recovery of the ozone hole over Antarctica and no returning to the 1980 level over the tropics by 2100. The results also strikingly demonstrate that both lower-stratospheric ozone and temperature are controlled by cosmic rays and ozone-depleting substances only. This study greatly improves quantitative understanding of ozone depletion and climate in the global lower stratosphere and offers new predictions on future trends.

**Introduction**

The mechanisms of ozone depletion in the stratosphere have been intensely studied over the past five decades (*1-4*). However, there remain surprising, unexpected observations from the conventional photochemical models. The recent examples include the observations of the largest ozone hole over the Arctic in 2020 (*5*) and the maximum and most persistent Antarctic $O_3$ holes (and associated maximum stratospheric cooling) in 2020 and 2021 (*6*), in striking contrast to the declining total level of surface (tropospheric)-measured ozone-depleting substances (ODSs) (mainly chlorofluorocarbons – CFCs) since the mid-1990s due to the successful execution of the Montreal Protocol and its Amendments. Over this long period, the status of ozone research remains essentially unchanged: although the chemistry-climate models (CCMs), which are based on the photochemical effects of ODSs on $O_3$, reproduced reasonably well the observed middle- and upper-stratospheric $O_3$ loss at northern mid-latitudes, there exist persistent large discrepancies between CCMs and observations of the largest ozone loss ('ozone hole') in the lower stratosphere (*2-4, 7-15*), indicating significant gaps in understanding of the partitioning of chlorine therein. The uncertainties in $O_3$ profile of pre-1997 $O_3$ trends in the lowermost stratosphere from both CCMs and ground- and satellite-based observations are very large with the uncertainties of $O_3$ loss up to ±27% per decade (see, for example, Figs. 5.9 and 5.10 of the 2019 SPARC's LOTUS Report (*3*) and Table 1 of ref. (*15*)). These discrepancies and uncertainties pose great challenges in ozone trend analysis.

In contrast, the above-mentioned observations are consistent with the prediction from our cosmic-ray-driven electron-induced-reaction (CRE) mechanism of ozone depletion in the Earth's atmosphere under cosmic ray (CR) ionization (*16-18*), which has been strongly supported by mounting observations over the past 4 solar cycles (*6, 9-11, 19*). For example, the prediction to observe two of the recent largest Antarctic holes in 2008 and 2020 was explicitly stated in the abstract of the PRL paper submitted in August 2008 by the author (*18*). Moreover, the fingerprints of the CRE mechanism were found in the spatial and temporal patterns of lower-stratospheric ozone and cooling at Syowa (69°S, 39.6°E), Antarctica for the period 1980-2020 (*6*). Others have also strong interest in studying the effects of CRs on the Earth's ozone layer and space weather (*20-24*), and our work (*16-18*) has stimulated continued interest in studying the relevant dissociative electron attachment/transfer (DEA and DET) reactions of ODSs, particularly CFCs (*25-32*) and hydrochlorofluorocarbons (HCFCs) (*33*), including those on clusters and ice surfaces with relevance to atmospheric and interstellar processes (*30, 32, 33*).

As noted by Randel et al.(*7*), the vertical profiles of $O_3$ trends can provide a fingerprint for the mechanisms of $O_3$ depletion. It is therefore crucial to obtain quantitative calculations of vertical profiles of $O_3$ loss and to compare them with observations. Fortunately the author (*19*) has recently formulated the CRE mechanism to provide a superior capability of making *no-parameter* quantitative calculations of global $O_3$ depletion; the CRE theoretical results have shown excellent agreement with the observed vertical profiles of *decadal-mean* $O_3$ loss from both ozonesonde and satellite measurements in the period 1980s (1960s) to 2000s.

Furthermore, the temporary variation of *lower-stratospheric ozone* (LSO) can not only provide another key fingerprint for the mechanisms of $O_3$ depletion but also plays an important role in climate models to capture the atmospheric response to solar cycle variability and to test the predicted effects of well-mixed non-halogen greenhouse gases (GHGs) ($CO_2$, $N_2O$ and $CH_4$) on the ozone layer. Simulations by $CO_2$-based climate models have long predicted that the greenhouse effect of increasing atmospheric concentrations of these non-halogen GHGs would largely enhance stratospheric ozone loss and associated stratospheric cooling (*34-37*). In striking contrast to this prediction, the global mean *lower-stratospheric temperature* (LST) has become rather stable since the mid-1990s(*2, 38, 39*), which is essentially controlled by the total level of ODSs. It is therefore important to quantify the magnitudes and patterns of the temporary variations of LSO and LST to understand the impacts of GHGs and the 11-year solar/CR cycle on the atmosphere and climate.

Maycock et al. (*40*) made multiple linear regression analyses to extract the so-called "solar cycle ozone response" (SOR) in satellite ozone datasets in the period 1970 to 2013. Their results yielded annual mean SOR amplitudes of ~1% and 2% in the tropical upper stratosphere in the SAGE II version 7.0 (v7.0) mixing ratio dataset (1984–2004) and the SBUV Merged Ozone Dataset (SBUVMOD) version 8.6 (VN8.6) (1970–2012) respectively. CCMs only address the impact of changes in incoming solar ultraviolet radiation over the 11-year solar cycle on stratospheric ozone abundances. Namely, a variation in solar UV radiation leads to a variation in $O_3$ production via the photolysis of $O_2$ in the *upper* stratosphere. This explanation does not involve any photochemical processes related to ODSs. Recent CCMs in the CMIP6 gave a peak amplitude of 2 % in the annual mean 11-year SOR in the *tropical upper* stratosphere (1–5 hPa) but no large (>2%) amplitudes in 11-year cyclic variations of polar or mid-latitude LSO, as reviewed by Maycock et al. (*41*). Most new-generation CCMs have not explicitly included any chemical effects of CRs and solar energetic particles, as is the case in global climate models that must prescribe an ozone dataset including a representation of the SOR (*41*). Attempting to include the solar effect in modeling the polar $O_3$ loss within photochemical mechanisms, atmospheric researchers conceded that it is difficult to understand the $O_3$ variation and predict future trends and that understanding and constraining the SOR is a longstanding scientific challenge (*41*).

It is also well-known that cooling in LST is a direct indicator of loss in LSO. In 1996, Ramaswamy et al. (*42*) investigated the fingerprint of ozone depletion in the spatial and temporal pattern of global LST cooling for the period of one solar cycle (1979-1990). Ten year later, Ramaswamy et al. (*43*) extended their studies for over two solar cycles (1979-2003). Unfortunately, they attributed the non-monotonic decrease in LST to the effect of volcanic eruptions without an 11-year periodicity.

In contrast to the above-mentioned studies based on the photochemical mechanisms of ozone depletion, pronounced (large, >2%) 11-year cyclic variations of LSO loss and associated LST cooling were uniquely predicted by the CRE mechanism (*17, 18*). Indeed, such a prediction was long criticized by some atmospheric researchers, such as Patra and Santhanam (*44*) and Müller (and Grooß) (*45-49*). Despite these rejections, the present author has shown robust observations of 11-year cyclic variations of LSO and LST in the Antarctic $O_3$ hole, corresponding to the 11-year cyclic variation of the CR intensity in the past 3-4 solar/CR cycles from 1979 to 2013/2020 (*6, 9-11*). Notably, the LST also exhibits an excellent linear correlation with measured total ozone at the Antarctic stations with a high linear correlation coefficient up to 0.94. The observed 11-year cyclic variations of LSO and LST were nicely fitted with an old-version CRE equation that had two inputs of the CR flux and the equivalent effective chlorine in the stratosphere plus a single fitting parameter (*6, 9-11*). Moreover, the present author also showed no apparent 11-year cyclic variations of ozone loss in the upper stratosphere at 1-2 hPa or of LSO or LST in the pre(no)-ozone hole season (e.g., winter or summer) over Antarctica, where or when the CRE reactions of ODSs are ineffective (*6, 9-11*). In fact, the above CRE prediction has now been confirmed by others (*50*). In the latter, "a large solar cycle signal (SCS)" in the Antarctic LSO and a SCS up to 6% in the Southern Hemisphere (SH) mid-latitude LSO were reported. In simulations using chemistry transport models (CTMs) in that study (*50*), however, no physical or chemical mechanism of ODS-related LSO depletion was actually included. Notably, CTMs require the use of historical meteorology such as measured stratospheric temperatures and winds (polar vortex dynamics) as model input parameters to produce the desirable outputs. Such requirements not only limit CTMs' ability to predict future changes of the $O_3$ hole (*51*) but lead to CTMs' intrinsic problems as stratospheric cooling is a direct consequence of $O_3$ loss. As a matter of fact, the observations of pronounced 11-year cyclic oscillations of LSO and LST in the Antarctic $O_3$ hole are indicative of the CRE being the dominant mechanism for the Antarctic ozone hole (*6, 9-11, 19*).

This article provides a quantitative understanding of LSO depletion in global regions and reports new results in three major aspects. First, using the formulated CRE equation (*19*), no-parameter theoretical calculations of the vertical profile of $O_3$ loss in the 2020 largest Antarctic hole are performed and the results are compared with the measurements of LSO and LST at the Antarctic Syowa station by the Japanese Meteorological Agency (JMA). Second, results from CRE theoretical calculations of time-series LSO variations over Antarctica, mid-latitudes and the tropics at various altitudes are presented and compared with LSO and LST observations from the available WOUDC ground-based ozonesonde and NASA satellite-based measured datasets. Third, CRE theoretical calculations offer the predictions on future changes in LSO and LST over Antarctica, mid-latitudes and the tropics under the three possible scenarios in future trend of CR fluxes and the baseline scenario of ODSs. These results should unravel the main mechanism for LSO depletion and associated LST cooling and render proper analyses of future measured data.

**Methods and Data**
As reviewed previously (*9-11*), abundant laboratory measurements have provided a solid physics and chemistry foundation for the CRE mechanism of ozone depletion (*16-18*), which has been strongly evidenced in substantial atmospheric observations. This mechanism comprises two major CR-driven processes, namely the charge-induced adsorption through a universal electrostatic bonding mechanism proposed by van Driel and co-workers (*52, 53*) and the subsequent DET reaction (*16, 17, 25, 27, 28, 54-56*) of ODSs on surfaces of atmospheric particles. The DET reactions of ODSs produce reactive neutral radicals and halogen anions and most halogen anions are trapped at ice surfaces due to the image potential. Furthermore, ionic reactions on surfaces can effectively convert $Cl^-/Br^-$ into photoactive $XNO_2$, $X_2$, and HOX (X=Cl, Br), as was demonstrated on aqueous sea-salt particle surfaces by Finlayson-Pitts and co-workers(*57, 58*). The primary conversion pathway is the reaction of $Cl^-$ with the nocturnal $NO_X$ species, $N_2O_5$, producing the photolabile nitryl chloride ($ClNO_2$) (*59*). Upon photolysis in the upper troposphere and lower stratosphere, the photoactive halogen species ($XNO_2$, $X_2$, and HOX) rapidly liberate reactive Cl/Br radicals. Additionally, the author(*19*) has quantitatively unraveled a new process that the denitrification (reduced $N_2O_5$) and associated anion conversion reaction also play an important role in controlling the lifetime of adsorbed $Cl^-$ on the surfaces and therefore surface charging and adsorption on atmospheric particles.

In the formulated CRE theory of ozone depletion (*19*), the average diurnal sum concentration of Cl atoms yielded from the CRE mechanism of all ODSs, expressed as the volumetric concentration by the particle surface area density μ in the atmosphere, is given by

$$[Cl] = \sum_i k^i \theta_{ODS}^i \Phi_e^2, \qquad (1)$$

where $\Phi_e$ is the flux of the prehydrated electron ($e_{pre}^-$) at the surface of the cloud or aerosol particles, $\theta_{ODS}^i$ is the coverage of the ODS species on the surface determined from Langmuir's adsorption equation, and $k^i$ is the effective (volumetric) dissociative electron transfer (DET) coefficient of an ODS (adsorbed on the particle surface) in the atmosphere. $k^i \equiv \mu k_e^i \tau_e \tau_{ion}$ is the product of the atmospheric particle surface area density µ, the DET reaction cross section $k_e^i$ for an adsorbed ODS species with an $e_{pre}^-$, and the lifetimes of $e_{pre}^-$ and Cl$^-$ ions trapped on the surface ($\tau_e$ and $\tau_{ion}$) (*19*).

In essence, O$_3$ is depleted by the reactive halogen atom in the catalytic reaction cycles leading to O$_3$ depletion, and the rate of O$_3$ loss can be expressed as its direct dependence on Cl and Br atomic concentrations [Cl] and [Br] (*60*)

$$-(d[O_3]/dt) \equiv (k_{Cl}[Cl] + k_{Br}[Br])[O_3] \qquad (2)$$

where $k_{Cl}$ ($k_{Br}$) is the rate constant for the reaction of Cl (Br) atom with ozone, with $k_{Cl}$ =2.9×10$^{-11}$exp(–260/T) cm$^3$s$^{-1}$ (*8*). Notably, Eq. 2 is equivalent to the expression of the O$_3$ depletion rate in terms of the ClO and BrO concentrations that are more readily detected than [Cl] and [Br] in the stratosphere (*61, 62*). However, Eq. 2 has an apparent advantage, which avoids the coupling term between [ClO] and [BrO]. It is convenient to use Eq. 2 to obtain the rate of stratospheric ozone loss in percentage.

Integration of Eq. 2 over a 24-hour period yields the amount of O$_3$ removed in each diurnal cycle (*60*). The halogen chemistry in the troposphere may be more complex as halogen atomic radicals may involve in the processes leading to not only ozone loss but also increased ozone production, especially in polluted regions (*59, 63, 64*). Despite this complexity, Eq. 2 has been used to study ozone depletion during the 'polar sunrise', i.e., the destruction of tropospheric ozone correlated with high average diurnal Cl and Br atom concentrations in polar regions(*60*) and mid-latitudes(*65*). Moreover, the atmospheric abundances of anthropogenic CFCs and HCFCs have been dominant in all ODSs, and there have been very small changes in concentrations of naturally generated CH$_3$Cl and Br-containing ODSs (mainly CH$_3$Br) since the 1960s. Additionally, the rate constant $k_{Cl}$ is 10 times larger than $k_{Br}$ at 298 K (*8, 60*). Therefore, time-series changes in *anthropogenic* ozone loss rate can be expressed as

$$-(d[O_3]/dt) \approx k_{Cl}[Cl][O_3] \qquad (3)$$

As demonstrated in our recent paper (*19*), this equation is used for calculating vertical profiles of O$_3$ trends in global regions, in which [Cl] is given by Eq. 1, in the current study.

In the stratosphere and troposphere, the primary source to produce electrons is atmospheric ionization by CRs originating from deep space. Upon entry into the Earth's atmosphere, CRs initiate nuclear-electromagnetic cascades, resulting in a maximum in ionization rate at the altitude 15-18 km, referred to as Pfotzer maximum (*21, 66, 67*). Owing to the geomagnetic field, CRs also exhibit a latitude-dependent effect: only CRs with energies exceeding 12 GeV are able to penetrate the tropical region while CRs of all energies impact the polar regions. Therefore, the ionization intensity of CRs in the polar stratosphere is much larger than that over the tropics (*66-68*). Besides, lower-energy CRs are susceptible to the impact of solar winds. Therefore, the CR flux is anti-phased with solar intensity with an average periodicity of 11 years (ranging in 9-12 years): when solar activity is at maximum, the CR intensity is at minimum, and vice versa. This 11-year cyclic modulation decreases with decreasing latitudes, being most pronounced in polar regions and the minimum over the tropics (*66-68*).

This study uses the following datasets: the high-quality monthly or daily altitude profiles of $O_3$ at Syowa (69°S, 39.6°E), Antarctica, obtained from the Umkehr and Ozonesonde datasets (*69*) at the Japan Meteorological Agency (JMA) website and from the World Ozone and Ultraviolet Radiation Data Centre (WOUDC); the monthly altitude profiles of global ozone depletion were obtained from the ground-based WOUDC ozonesonde Trajectory-mapped Ozonesonde dataset for the Stratosphere and Troposphere (TOST_SM) (*70*) and the NASA satellite (GOZCARDS _source and _merged) (*71*) datasets; the monthly or annual LSTs at individual Antarctic stations and global regions were obtained from the NOAA RATPAC-B datasets; surface-measured and projected atmospheric concentrations of ODSs were obtained from the baseline scenario given in the 2022 WMO Ozone Assessment Report (*4*), and lags of zero, five and ten years are applied to derive the lower-stratospheric concentrations of ODSs in the Antarctic, mid-latitudes and tropics respectively, where the lags are estimated from our observed LSO trends (*6, 10, 11*); stratospheric CR fluxes at the maximum of the absorption curve in the atmosphere for the period 1957–2022 were obtained from the Laboratory of Physics of the Sun and Cosmic Rays, P.N. Lebedev Physical Institute of the Russian Academy of Sciences and were updated to 2024 by a correlation R=1 to the surface-measured counts from Neutron Monitor datasets, and the altitude profiles of CR fluxes in the Antarctic, mid-latitudes and tropics were obtained from Bazilevskaya et al (*21*). TOST is a global 3-D (latitude, longitude, altitude) climatology of tropospheric and stratospheric $O_3$, derived from the WOUDC global $O_3$ sounding record by trajectory mapping and uses approximately 77,000 ozonesonde profiles from some 116 stations worldwide since 1965, with details given by Liu et al (*70*). The NASA GOZCARDS is a combination of various high-quality space-based monthly zonal mean ozone profile data, with details given by Froidevaux et al (*71*). Among long-term merged satellite data sets (SBUV MOD/COH, GOZCARDS, SWOOSH, SAGE-OSIRIS/CCI/MIPAS-OMPS) relying mostly on NASA's SAGE I/II data for pre-2000 periods, GOZCARDS is the only merged satellite dataset that extends to the lowermost stratosphere and covers the early satellite era starting in 1979, as reviewed in the recent LOTUS Report (*3*).

**Results and Discussion**
Fig. 1A plots daily altitude profiles of ozone at Syowa (69°S, 39.6°E) in the pre-ozone depletion season (March, April to May, MAM) and the deepest ozone-hole month (October) in one of the recent deepest Antarctic $O_3$ holes observed in 2020, which set a lowest record of 128 DU at Syowa on 3rd October 2020 since the 1960s (*6*). As noted previously (*6*), the $O_3$ layer at 14-18 km, at which the peak of the CR ionization rate is located, has constantly been completely depleted in the largest and deepest $O_3$ holes typically appearing at the years of CR peaks during 11-year cycles since the late 1990s. In this lower stratospheric layer of ~4 km, the $O_3$ layer is less sensitive to the CR maxima as the CRE reactions of ODSs have over-killed all the $O_3$ molecules. Outside this stratospheric layer, the column $O_3$ has only been partially depleted and therefore $O_3$ depletion is more sensitive to the cyclic CR maxima. Fig. 1B plots the observed seasonal or monthly mean altitude profiles of ozone for the pre-ozone depletion season (MAM) and the deepest ozone hole month (October) and the thus measured altitude-profile of October monthly mean $O_3$ loss in percentage in the $O_3$-hole period relative to the pre-$O_3$ depletion season of 2020 at Syowa, in comparison with the CRE theoretical profile of $O_3$ loss calculated by Eqs. 1 and 3. The results exhibit a surprising perfect agreement between observations and calculations, particularly in the altitude range 12-25 km where the undisturbed ozone layer is thick, and both measured and calculated ozone profiles give an identical ozone loss maximum of –97.7% at ~17 km. Given the fact that no adjustable parameters are included in CRE theoretical calculations (*19*), this perfect agreement is very remarkable, indicating the super capability of the CRE theory to provide a precise understanding of LSO depletion over Antarctica. In the troposphere below 10 km, theoretical ozone losses are somewhat larger than observed data, which can reasonably be explained by the enhanced ozone production due to air pollution at Syowa. This is slightly different from the average case for the entire Antarctic (60°-90° S), which is essentially pollution free (*38, 39*). For the latter, our calculated results are excellent agreement with observations, as shown previously (*19*). Below 5 km or above 30 km where the CRE reactions of ODSs are ineffective, theoretical results of ozone depletion by the CRE mechanism become negligibly small. This is also in excellent agreement with observed data, which also show no visible 11-year cyclic variations (*6*). Overall, the results in Fig. 1 clearly demonstrate the unprecedented capability of the CRE theory to give accurate calculations of ozone depletion in the Antarctic lower stratosphere.

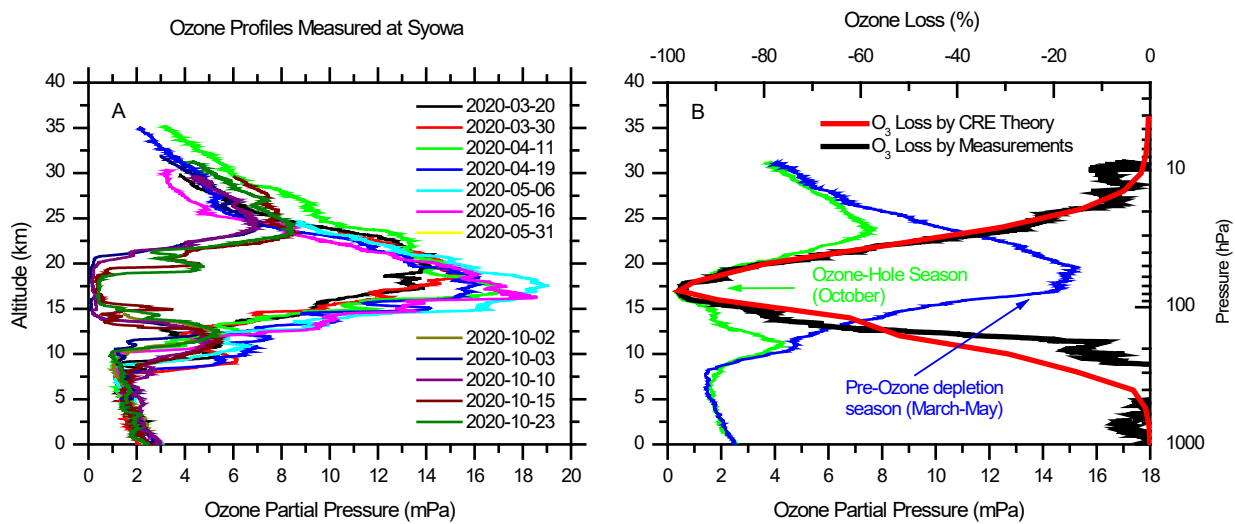

**Fig. 1. A**. Measured daily altitude profiles of ozone in the pre-ozone depletion season (March-May, MAM) and the deepest Antarctic ozone-hole month (October) of 2020 at Syowa (69°S, 39.6°E). **B**. The observed mean altitude profiles of ozone (thin blue and green lines) for the MAM season and October shown in A, the altitude profile of October monthly mean $O_3$ loss in percentage (thick black line), and the altitude profile of ozone loss calculated by the CRE theory (thick red line).

Fig. 2A shows observed time-series October monthly mean $O_3$ loss at various stratospheric layers (127-63, 63-32, 32-16, and 16-8 hPa) and October monthly mean stratospheric temperatures at 100, 50 and 30 hPa at Syowa as well as the CRE theoretical $O_3$ loss in the period 1979-2023. For a quantitative comparison, an inevitable difficulty is to find the initial (undisturbed) ozone amounts presumably in the 1950s-1960s when there lacked reliable measurements. In this study, the initial ozone amounts for all the observed results of ozone loss presented below are chosen to match the overall theoretical curves. The results in Fig. 2A provide important information to shed light on the mechanisms of $O_3$ depletion. First, both the observed unpredicted dips in LSO and LST in 2015 and peaks in 2019 were reasonably attributed to the well-documented volcanic eruption of Calbuco in 2015 and a sudden stratospheric warming in 2019 respectively, and both effects became diminished at higher altitudes ≤ 30 hPa, as observed previously (*6*). Second, pronounced 11-year cyclic variations in both ozone and temperature at the altitude range of 100-8 hPa are clearly seen over the past 3-4 solar/CR cycles. Third, time-series observed data of ozone and temperature including marked 11-year cyclic variations are excellently reproduced by no-parameter CRE theoretical calculations, which also directly give both ozone loss and associated stratospheric cooling maxima around 2008 and 2020. The overall oscillatory amplitudes up to ~20% in calculated $O_3$ loss are slightly larger than those in observed data at various altitudes, whereas the peaks (minima) of calculated ozone loss are generally smaller than those of observed ones that are up to 25-30%. Fourth, the over-killing effect of the CR-driven reaction at the CR ionizing peak region (127-63 hPa or 100 hPa) is confirmed in observed data, leading to a smaller 11-year oscillatory amplitude. Finally, no apparent 11-year cyclic ozone variations are seen in the upper stratosphere (2-1 hPa), as shown previously (*6*). This is also consistent with the CRE calculations as the CRE reactions of ODSs are ineffective in the upper stratosphere (see Fig. 1).

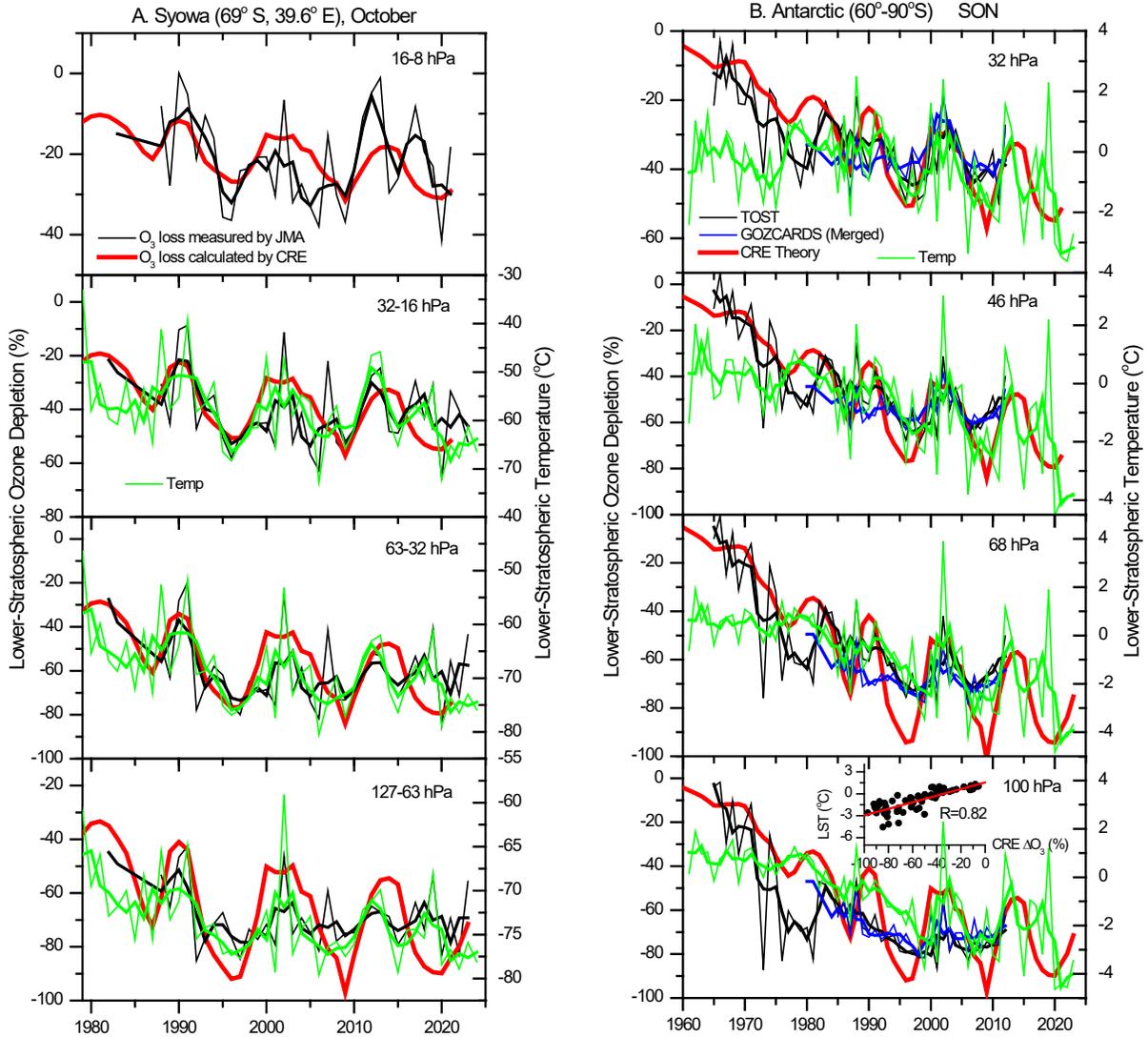

**Fig. 2. A.** Time-series observed October monthly mean ozone loss in percentage at various stratospheric layers (127-63, 63-32, 32-16, and 16-8 hPa) and October monthly temperatures at various altitudes of 100, 50 and 30 hPa at the Antarctic Syowa station (69°S, 39.6°E) as well as theoretical ozone loss calculated by the CRE theory (thick red line) during 1982-2023. **B.** Time-series observed September to November (SON) mean ozone loss in percentage at various stratospheric layers (100, 68, 46, and 32 hPa) and annual temperature at 100, 70, 50, and 30 hPa in the Antarctic (60°S–90°S) as well as theoretical ozone loss calculated by the CRE theory (thick red line) during 1960-2023. The observed ozone data were obtained from the WOUDC TOST (-SM) (black line) and NASA GOZCARDS (Merged) (blue) datasets and temperature (green) data were from the NOAA RATPAC-B datasets. The right y-axis ranges for temperatures (green lines) in both A and B are chosen to best match ozone data. For a guide to the eye, a minimal 3-year smoothing (thick lines) is applied to observed data (thin lines) for both temperature and ozone. The inset in B for 100 hPa shows a good linear correlation between observed LST (3-year smoothed) and CRE calculated LSO loss with a correlation coefficient R=0.82.

Fig. 2B shows observed results for the entire Antarctic (60°S–90°S) in the spring season (September-November, SON) from the WOUDC TOST_SM dataset starting in the 1960s and NASA GOZCARDS dataset starting in 1979, together with CRE theoretical results. These results are very similar to those presented in Fig. 2A. Note that compared with Fig. 2A, only annual mean Antarctic LST data are plotted in Fig. 2B since no October monthly or SON mean Antarctic LST data are available in the RATPAC dataset. However, we have previously demonstrated that both the Antarctic ozone depletion and associated stratospheric cooling mainly occur in the spring season only, and therefore the annual mean LSTs have varying trends closely similar to those in October monthly or SON, as seen in Figs. 2A and 2B. The results also show that the observed LST data have more pronounced 11-year cyclic variations and are in better agreement with CRE theoretical results than the observed LSO data, except for the period 1960s–1970s when there lacked reliable measurements. This is probably due to the larger uncertainties in measured ozone data. As we expected, there is a good linear correlation between observed LST change and CRE-calculated Antarctic ozone loss with a correlation coefficient R=0.82, as shown in the inset in Fig. 2B for the altitude 100 hPa. Overall, the results in Figs. 2A and 2B demonstrate the robustness of the observations of 11-year cyclic variations in both Antarctic LSO and LST and their quantitative agreement with CRE theoretical calculations. Another important result is that in contrast to the significant and continuous decreases of the concentrations of major ODSs since the mid-1990s, both original observed data of LSO and LST and CRE calculations show no clear recovery in the Antarctic. This delay is due to the effect of an increased CR trend over the past few solar cycles. As we demonstrated previously, a pronounced recovery in LSO can be revealed either if the effect of the increased CR trend is removed from observed springtime ozone data or if ozone data in summertime when the CRE reactions of ODSs are ineffective are analyzed (*6, 10, 11*). The effects of possible CR future trends on the Antarctic ozone trend will be further demonstrated by CRE calculations given later in this paper.

Figs. 3A and 3B show observed and theoretical results of LSO and LST for the Southern Hemisphere (SH) mid-latitudes (30°S–60°S) and Northern Hemisphere (NH) mid-latitudes (30°N–60°N) respectively. Here two important features are highlighted. First, the CRE theoretical calculations give ozone loss of 8.3% and 8.6% per decade on the average at 15 km (100 hPa) altitude in the period 1980-2000 for the SH and NH mid-latitudes respectively, which are in good agreement with the widely-accepted value of 7.3 ± 4.6% per decade during 1979-1997 at 15 km altitude of NH mid-latitudes where most balloon and ground-based measurements are made (*7*). Second, the calculations also give an 11-year oscillation amplitude of 4.5% for NH and 4.3% for SH in mid-latitude LSO since 1990, which roughly agree with the observed data though the latter have large uncertainties (are very noisy). Overall, the theoretical results are in good agreement with observations at mid-latitudes. Additionally, there is a weak recovery trend in LSO at mid-latitudes, which is somewhat more significant at the SH.

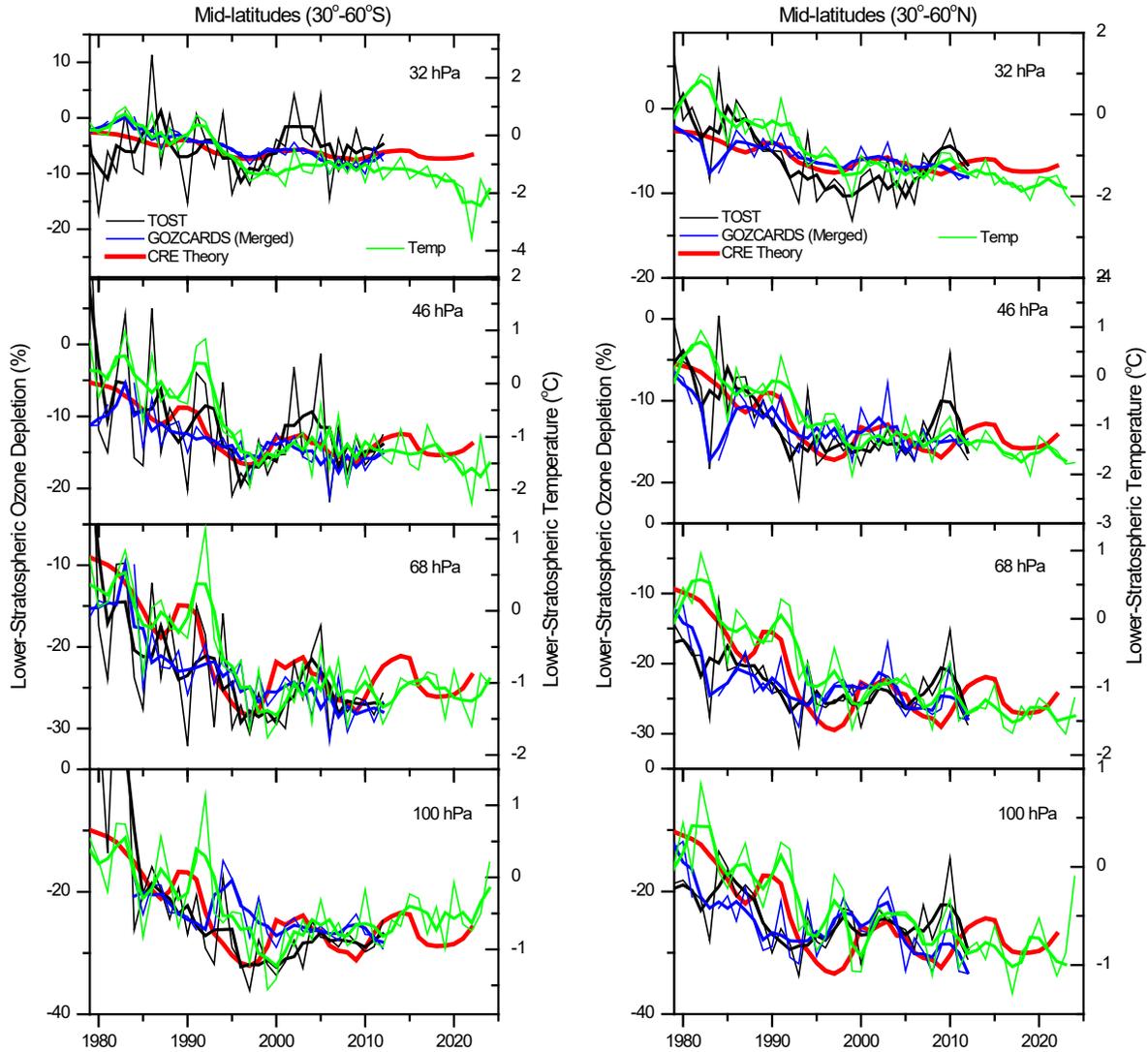

**Fig. 3. A-B:** Time-series data of observed annual mean ozone loss in percentage and temperature at various stratospheric layers (100, 68, 46, and 32 hPa) in SH mid-latitude (30°S–60°S) and NH mid-latitude (30°N–60°N) as well as theoretical ozone loss calculated by the CRE theory (thick red line) during 1979-2023. The observed ozone data were obtained from the WOUDC TOST (-SM) (black line) and NASA GOZCARDS (_Merged) (blue line) datasets and temperature (green line) data were from the NOAA RATPAC-B datasets. For a guide to the eye, a 3-year smoothing (thick lines) is applied to observed data (thin lines).

Fig. 4 shows observed and theoretical results of LSO and LST for the tropics (30°S–30°N). In view of the facts that there exist large uncertainties and strong controversies about the tropical LSO depletion in the literature (*7, 15, 72, 73*), the results in Fig. 4 deserve for special discussion. At the tropical lowermost stratospheric altitude 100 hPa, the CRE theoretical calculations give a maximum ozone loss of −75.5% in the 2000s, which is in excellent agreement with both the TOST_SM and GOZCARDS_Source data since 1979. The GOZCARDS_Merged dataset gives a smaller ozone loss by about 10%, which arises likely from the data merging processing from SAGE-I to SAGE-II. Both observed and calculated results consistently show a gradual recovery of the tropical LSO layer since around 2005. At 68 hPa and higher altitudes, there are large discrepancies between the CRE calculations and TOST_SM data (not shown here) (*19*), whereas the CRE calculations show excellent agreement with both GOZCARDS_Source (not shown) and GOZCARDS_Merged data, especially at 46 and 32 hPa. As noted previously, there are few ground-based observation stations in the tropical region and thus larger uncertainties in TOST_SM data (*19*). A striking feature for the tropical LSO and LST is the absence of apparent 11-year cyclic variations in both observed data and CRE theoretical calculations using measured CR fluxes. When ozone depletion was qualitatively estimated from the CRE mechanism, we expected to observe 11-year cyclic variations in tropical LSO as the LSO varying amplitude has a quadratic dependence on the CR flux (*72*), as implied in Eq. 1. Given the uncertainties (noisy levels) of both measured ozone and CR data, however, our current no-parameter CRE theoretical calculations give rise to a result that the oscillatory amplitude of the tropical LSO loss is barely visible (too small to see). This is because in the tropical lower stratosphere, the CRs are far more energetic (harder) and less affected by solar winds, and therefore the CR flux has a 11-year variation amplitude (~5%) far less than that in the polar lower stratosphere (~20%) (shown below). The uncertainties in both measured ozone and CR fluxes blur the expected small 11-year cyclic oscillations in tropical ozone loss. In marked contrast to this characteristic of the CRE mechanism, the CCMs have predicted to observe most significant solar (11-year) cyclic variations of ozone in the tropical upper stratosphere but none in the polar lower stratosphere (*40, 41*). The overall good agreement between the CRE calculations and observed data shown in Fig. 4, together with Figs. 1-3, strongly indicates that the previous observations of a large (largest) ozone loss up to 80% in the lowermost stratosphere over the tropics (*7, 15, 19, 72, 73*) are trustable and therefore robustly validated. Current no-parameter CRE calculations should greatly enhance the confidence in assessing and interpreting LSO loss in the tropics, which is long sought (*7*).

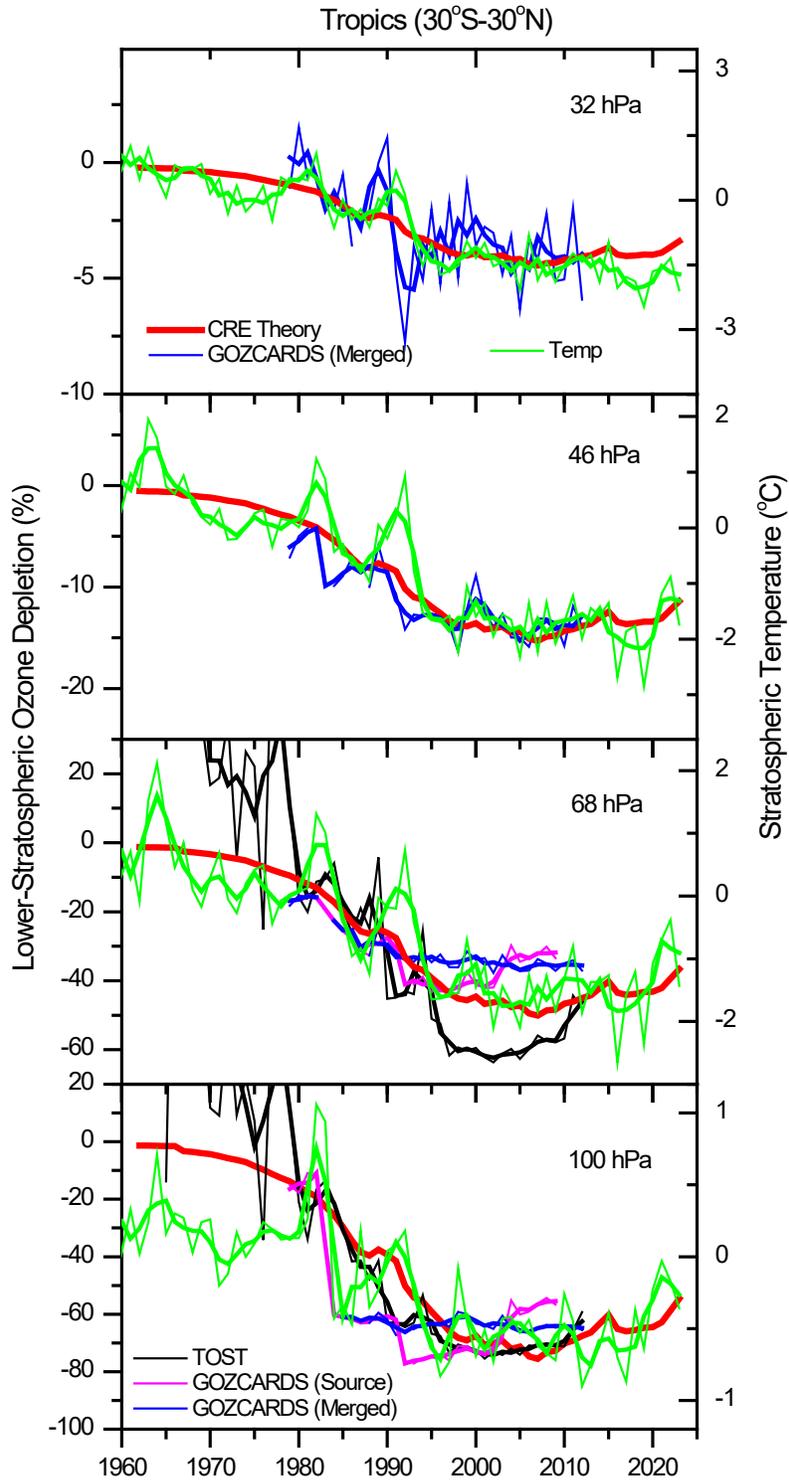

**Fig. 4.** Time-series data of observed annual mean ozone loss in percentage and temperature at various stratospheric layers (100, 68, 46, and 32 hPa) in the tropics (30°S–30°N) as well as theoretical ozone loss calculated by the CRE theory (thick red line) during 1960-2023. The observed ozone data were obtained from the WOUDC TOST (-SM) (black line) and NASA GOZCARDS (Source/Raw and Merged) (magenta and blue) datasets and temperature (green) data were from the NOAA RATPAC-B datasets. For a guide to the eye, a 3-year smoothing (thick lines) is applied to observed data (thin lines).

After obtaining quantitative agreement between observed and CRE theoretical calculations of LSO up to 2023 shown in Figs. 1-4, we present our calculations of future possible LSO trends in the Antarctic, mid-latitude and tropics with three projected CR scenarios and the baseline scenario of ODSs to the end of this century in Figs. 5 and 6, Figs. S1 and S2 in Supporting Materials (SM). The CR-intensity variation with an average periodicity of 11 years (9-12 years) and its modulation can approximately be expressed as (*9, 10*)

$$I_i = I_{i0}\left\{1 + \omega\,sin[\frac{2\pi}{11}(i - i_0)]\right\}, \qquad (4)$$

where $I_{i0}$ is the median CR intensity at year $i_0$ (e.g., 2005) in an 11-year cycle, and ω is the 11-year cyclic oscillation amplitude which is approximately 20%, 10% and 5% at the 15 km altitude for the polar, mid-latitudes, and tropics respectively, determined from the best fits to measured CR fluxes since the late 1950s. Note that the $I_{i0}$ values have kept increasing in the past four solar (CR) cycles, and the best fits to all the observed CR data from 1980-2024 yielded $I_{i0}$ increasing rates of ~4%, ~2% and ~0.3% per 11-year cycle for the polar, mid-latitude, and tropical low-stratospheric CR fluxes respectively (Fig. 5A). It is difficult to predict the future change of solar activity, which regulates the future change of CR fluxes in the coming decades. As presented previously (*9, 10*), however, we can construct three possible CR scenarios: A, the mean CR intensities ($I_{i0}$) in the coming solar cycles in the 21th century would keep the same rising rates as those in the past four solar cycles; B, $I_{i0}$ would keep nearly constant values identical to those observed for the past 11-year cycle (2013-2024); and C, $I_{i0}$ would reversely have decreasing trends returning to the values in 1980. The future CR intensities calculated by Eq. 4 with these scenarios of $I_{i0}$ are shown in Figs. 5 and 6, in which the measured and future concentrations of major ODSs (CFC-12 and -11) in the baseline scenario of ODSs given in the 2022 WMO Report (*4*) are also shown.

With the above projections of CR fluxes and ODSs, CRE theoretical calculations of October monthly $O_3$ loss at the 100, 68 and 46 hPa altitudes over Antarctica, NH mid-latitudes and the tropics in 1980-2100 by Eqs. 1, 3 and 4 are shown in Figs. 5 and 6, and Figs. S1, and S2 in SM.  Namely, Fig. 5 is for the CR scenario A, while Figs. 6, S2 and S3 are for CR scenarios B and C. These theoretical results of LSO show interesting complex behaviors that depend on the CR scenarios. If the CR flux follows the CR scenario A, strikingly the Antarctic ozone hole would have a very small recovery at 100 hPa and almost no recoveries at 68 and 46 hPa till the end of this century (2100), as shown in Figs. 5A and 5B respectively.

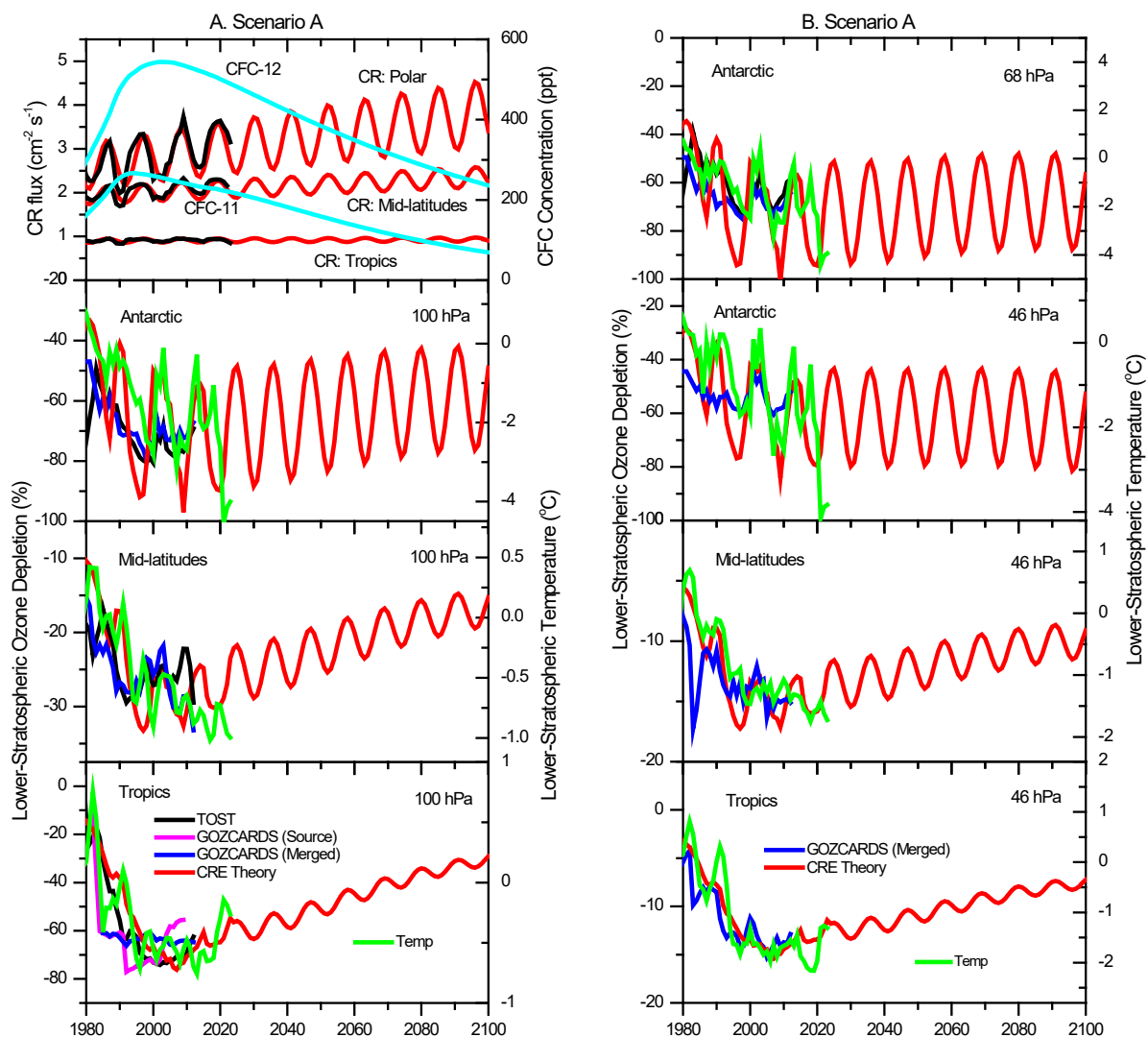

**Fig. 5.** Time-series observed annual mean ozone loss in percentage and temperature at various stratospheric altitudes (100, 68, and 46 hPa) in the Antarctic (60°S–90°S), NH mid-latitudes (30°N–60°N), and tropics (30°S–30°N) during 1980-2023 and theoretical ozone loss calculated by the CRE theory (thick red line) under the CR scenario A for the period 2024-2100. The observed and calculated ozone data for the period 1980-2023 are the same as Figs. 2-4.

The ozone layers at 100 and 46 hPa of mid-latitudes and the tropics would gradually recover but would not recover to their respective 1980 levels by 2100. In this scenario, the effect of the continuously increased CR fluxes obviously offsets that of declining trends in ODSs. For the CR scenario B, the Antarctic ozone hole would barely recover to the 1980 level by 2100, whereas the mid-latitude and tropical LSO layers would not recover to their 1980 levels by 2100. For the CR scenario C, the Antarctic and mid-latitude LSO layers would gradually recover to the 1980 level by around 2060 and 2100 respectively, whereas the tropical LSO layer would not recover to the 1980 level by 2100. It follows that in all the three CR scenarios, the tropical LSO layer have slowly recovered since the mid-2000s but will not recover to the 1980 level by the end of this century. Notably, the CRE calculated results of LSO exhibit small but clear 11-year cyclic variations in the tropics when the CR fluxes projected by Eq. 4 for the period 2024-2100 are used, in contrast to the CRE calculated results of LSO for the pre-2024 period when the measured CR fluxes are instead used. This clearly demonstrates that the uncertainties in CR flux and $O_3$ measurements can indeed smear the small oscillation amplitude of ozone in the tropical stratosphere, leading to its absence in both measured and calculated values of ozone loss.

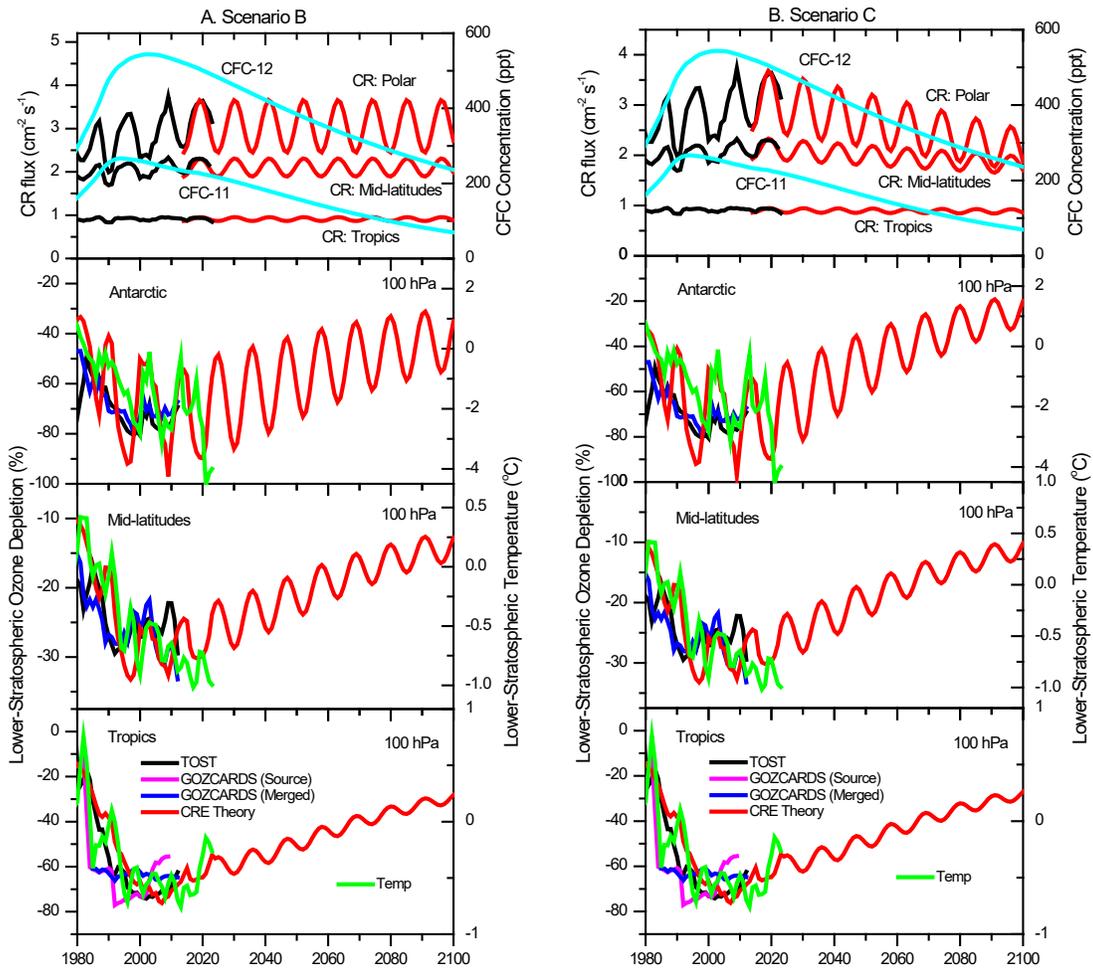

**Fig. 6.** Time-series observed annual mean ozone loss in percentage and temperature at the stratospheric altitude 100 hPa in the Antarctic (60°S–90°S), NH mid-latitudes (30°N–60°N), and tropics (30°S–30°N) during 1980-2023 and theoretical ozone loss calculated by the CRE theory (thick red line) under the CR scenarios B and C for the period 2024-2100. The observed and calculated ozone data for the period 1980-2023 are the same as Figs. 2-4.

The above CRE calculated results of LSO depletion with the three projected CR trends are very different from the projections by CCMs. The newest 2022 WMO Report (*4*) based on CCMs gives that the total column ozone would return to 1980 values around 2066 in the Antarctic, around 2045 in the Arctic, and around 2040 for the near-global average (60°N–60°S). Only the CRE result of the Antarctic LSO recovery under the CR scenario C is close to the CCM projection, whereas the CRE results for the Antarctic ozone hole under the CR scenarios A and B and for the mid-latitude and tropical LSO layers under any one of the three CR scenarios (A, B and C) are very different from the CCM projections.

In also marked contrast to the prediction of a gradual recovery in the tropical LSO layer since ~2005 from our CRE mechanism shown previously (*72*) and in Figs. 5, 6, S1, and S2, Chipperfield et al. (*74*) and Kuttippurath et al. (*75*) have argued that LSO decreases over the tropics are dominantly caused by dynamical processes, namely the strengthening of the BDC [i.e., increases in tropical upwelling and enhanced mixing between tropics and subtropics due to increasing well-mixed GHGs ($CO_2$, $N_2O$, and $CH_4$)] (*76-79*). It must be pointed out that such a dynamical mechanism can only cause small LSO decreases (approximately 2% per decade) over the tropics (*76-78*), which are approximately 10 times smaller than the observed negative LSO trends over the tropics in the 1980s and the 1990s (*15, 72, 73*), as also shown in Figure 4. Most notably, such a dynamical mechanism, through CCM simulations, predicted continuous *negative* (decreasing) LSO trends over the tropics till the end of this century (*74, 75, 77-79*), which are oppositive to our observed and CRE calculated results of LSO shown in Figures 4-6, S1, and S2. Future observations will determine which mechanism prevails.

**Conclusions**

Despite the significant and continuous decreases in atmospheric ODSs since the mid-1990s, there still lack clear-cutting recovery signs in LSO and LST. This study demonstrates that the CRE model has the superior capacity of providing a quantitative understanding of global ozone depletion and associated stratospheric cooling and can remove the persistent discrepancies between CCMs and observations, particularly in the lower stratosphere. No-parameter CRE theoretical calculations of LSO depletion show quantitative agreement with observations in the Antarctic, midlatitudes, and the tropics. Particularly, there is remarkably perfect agreement between the observed altitude profile of ozone loss in the deepest October Antarctic ozone hole at Syowa and CRE calculations. Furthermore, the results robustly confirm the observations of 11-year cyclic variations in both LSO and LST in the Antarctic and midlatitudes, whereas such cyclic variations are barely seen in the tropics. Moreover, these observed time series data of LSO loss in all the regions can be well reproduced by CRE calculations that use only two inputs, namely the concentrations of ODSs and the flux of CRs. The results also validate the recent discovery of large and all-season LSO loss in the tropics (the tropical ozone hole). Future trends of the CR fluxes can cause complex phenomena in changes of the Antarctic ozone hole. In the CR scenario that the CR fluxes keep the same rising rates as those in the past 4 solar cycles, there would strikingly be no recovery of the Antarctic ozone hole till the end of this century. Another surprising finding from CRE calculations is that the tropical LSO layer has slowly recovered since around 2005 but will not return to the 1980 level by 2100 under any one of the three projected CR scenarios.

The comparisons between observed results of LSO and LST in the past five decades and CRE calculated results of LSO strongly indicate that the predicted large greenhouse effect (climate forcing) of increasing non-halogen GHGs ($CO_2$, $N_2O$ and $CH_4$) in $CO_2$-based climate models on stratospheric ozone depletion and associated stratospheric cooling is missing, and instead, both LSO and LST are predominantly controlled by natural CR fluxes and anthropogenic ODSs. This is consistent with our previous observations and conclusion (*9-11, 19, 39*) and partially with some of the conclusions in the 2018 and 2022 WMO Reports (*2, 4*) and the IPCC Report AR6(*38*).

The results in this study emphasize the importance of carefully processing the ground- or satellite-based measured data to capture the real impacts of the natural CRs and anthropogenic ODSs on the atmosphere and climate. A cautious data processing is critical as both LSO and LST exhibit 11-year cyclic variations that were mostly ignored in previous ozone and climate research. The latter could readily cause large errors in merging different datasets measured in various periods such as SAGE I and II. Future trends of LSO and LST in observed data over the Antarctic and tropics, if recorded reliably and processed correctly, will decide which ozone depletion mechanism and even which global warming mechanism are dominant and prevail.

**Supporting Information (SI)**

See Supporting Information (SI) for Figures S1 and S2.

**Acknowledgments**

The author is greatly indebted to the Science Teams (JMA's Umkehr, WOUDC's TOST, NASA's GOZCARDS and HIRDLS, SPARC Data Initiative, EUMETSAT's ROM SAF, NOAA's RATPAC, etc.) for making the data used for this study available. This work is supported by the Natural Science and Engineering Research Council of Canada and the University of Waterloo.

**Data Availability**

The data used for this study were obtained from the following sources: the monthly or daily altitude profiles of $O_3$ at Syowa (69°S, 39.6°E), Antarctica, obtained from the Umkehr and Ozonesonde datasets (*69*) for the Japan Meteorological Agency (JMA) website (https://www.data.jma.go.jp/env/data/report/data/download/oznuv_e.html) and from the WMO's World Ozone and Ultraviolet Radiation Data Centre (WOUDC)( https://woudc.org/data/explore.php); the TOST data (*70*) were obtained from the WOUDC (https://woudc.org/archive/products/ozone/vertical-ozone-profile/ozonesonde/1.0/tost/); the GOZCARDS data (*71*) were obtained from the NASA EARTHDATA dataset (https://disc.gsfc.nasa.gov/datasets?keywords=GOZCARDS); the altitude profiles and DET cross sections of ODSs were obtained from refs. (*8, 80*) and refs. (*11, 16, 17, 55, 56, 81*), respectively; the RO lower stratospheric temperature satellite datasets were obtained from the ROM SAF (https://www.romsaf.org/product_archive.php); the monthly or annual LSTs at individual Antarctic stations and global regions were obtained from the NOAA RATPAC-B datasets (https://www.ncei.noaa.gov/pub/data/ratpac/); the CR flux data were obtained from ref. (*21*); surface-measured and projected data of ODSs were obtained from the 2022 WMO Report(*4*).

# SUPPLEMENTARY MATERIALS (SM)

# Observations and Theoretical Calculations of 11-Year Cyclic Variations in Lower-Stratospheric Ozone Depletion and Cooling

Qing-Bin Lu

Department of Physics and Astronomy and Departments of Biology and Chemistry, University of Waterloo, 200 University Avenue West, Waterloo, Ontario, Canada (Email: qblu@uwaterloo.ca)

The **Supplementary Materials (SM)** includes Figs. S1 and S2.

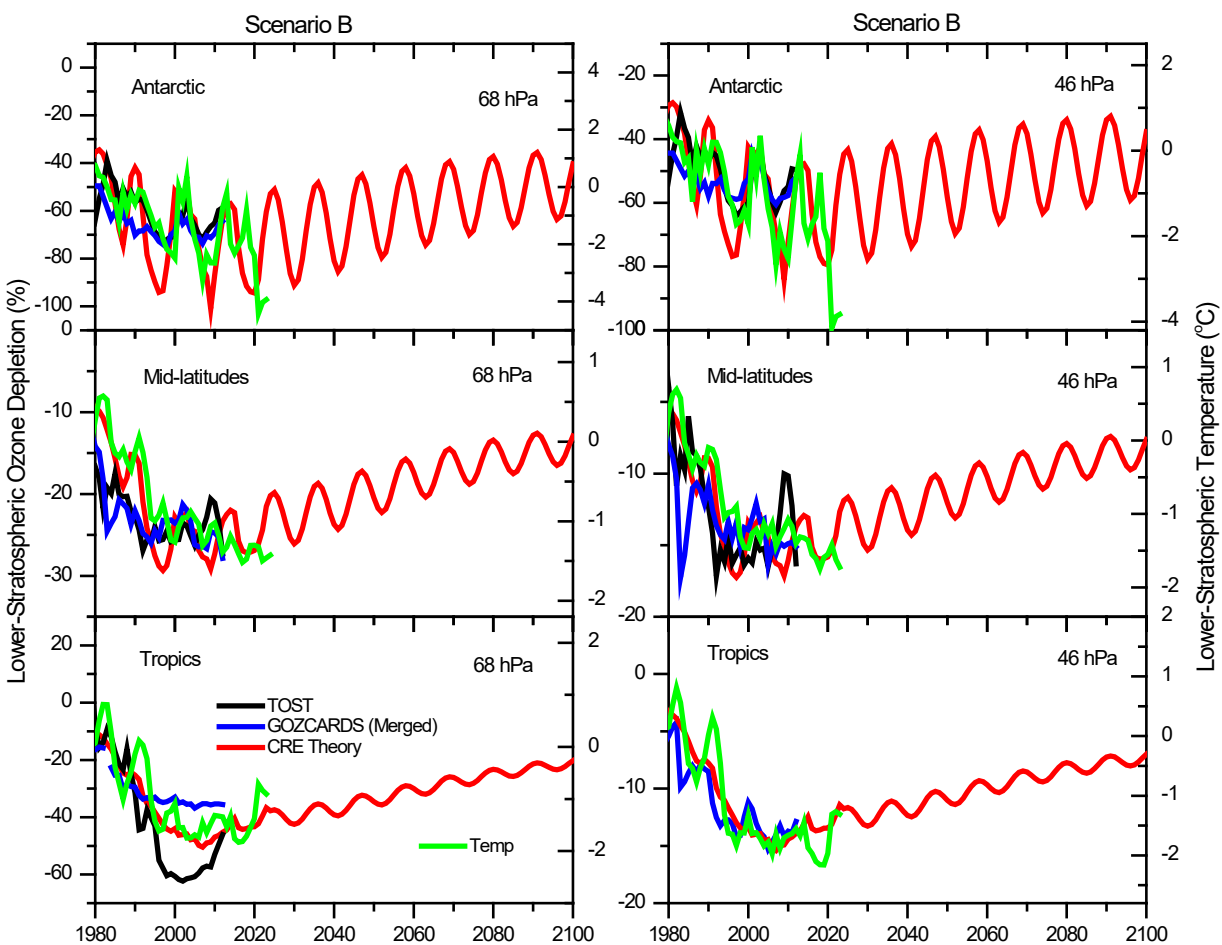

**Fig. S1.** Time-series observed annual mean ozone loss in percentage and temperature at the stratospheric altitudes 68 hPa and 46 hPa in the Antarctic (60°S–90°S), NH mid-latitudes (30°N–60°N), and tropics (30°S–30°N) during 1980-2023 and theoretical ozone loss calculated by the CRE theory (thick red line) under the CR scenario B for the period 2024-2100. The observed and calculated ozone data for the period 1980-2023 are the same as Figs. 2-4.

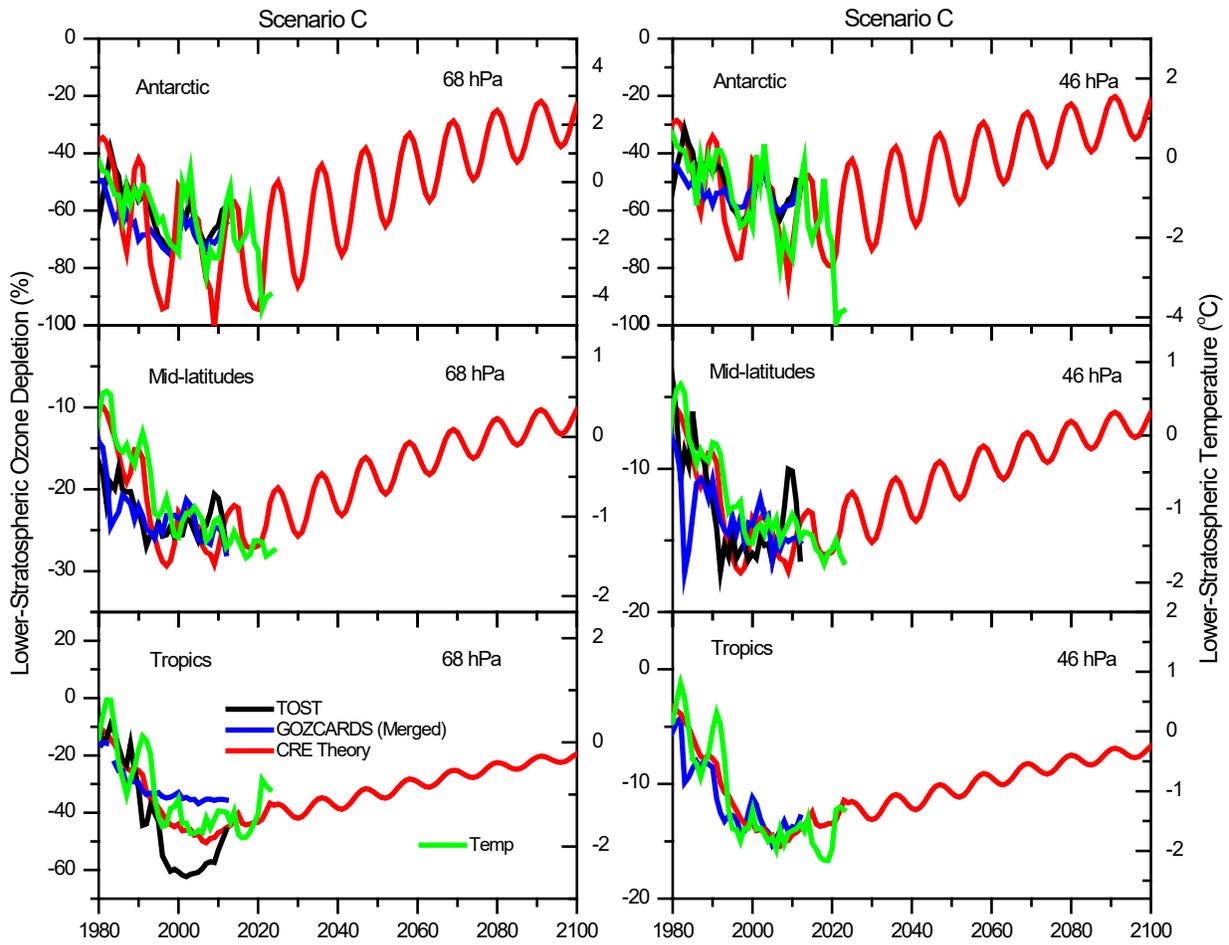

**Fig. S2.** Time-series observed annual mean ozone loss in percentage and temperature at the stratospheric altitudes 68 hPa and 46 hPa in the Antarctic (60°S–90°S), NH mid-latitudes (30°N–60°N), and tropics (30°S–30°N) during 1980-2023 and theoretical ozone loss calculated by the CRE theory (thick red line) under the CR scenario C for the period 2024-2100. The observed and calculated ozone data for the period 1980-2023 are the same as Figs. 2-4.